\documentstyle[12pt]{article}
\makeatletter
\gdef\@pubnumber{\null}
\long\def\pubnumber#1{\gdef\@pubnumber{CERN-TH/95-187\\DAMTP 95-14}}
\def\@makepub{\vbox to \z@{\hbox to
\textwidth{\hfill\llap{\parbox[t]{0.33\textwidth}{\raggedleft\@pubnumber}}}%
\vss}}
\def\@maketitle{\newpage
\@makepub \null
\vskip 2em \begin{center}
{\LARGE \@title \par} \vskip 1.5em {\large \lineskip 0.5em
\begin{tabular}[t]{c}\@author
\end{tabular}\par}
\vskip 1em {\large \@date} \end{center}
\par
\vskip 1.5em}
\pubnumber{1}

\def\ha{\frac{1}{2}}

\begin{document}

\setlength{\baselineskip}{5ex}

\title{The Stability of Primordial Magnetic Fields
  Produced by Phase Transitions}

\author{Adrian P. Martin$^1$ and Anne-Christine Davis$^{1,2,}$\\
        \\
        {\normalsize $^1$DAMTP, Cambridge University,}\\
        {\normalsize{Silver Street, Cambridge, CB3 9EW, UK}}\\
        \\
        {\normalsize $^2$Theory Division, CERN,}\\
        {\normalsize{Geneva 23, CH-1211, Switzerland}}}

\maketitle

\begin{abstract}
Primordial magnetic fields seem to be a generic relic of phase transitions
in the early universe. We consider a primordial electromagnetic field formed
as a result of a second-order phase transition, and show that it is stable to
thermal fluctuations in the period immediately following. We also
show how such
a field arises in first order phase transitions. In both cases there is a
transitive electric field produced during the transition.
\end{abstract}
\vskip 170pt
\flushleft {CERN-TH/95-187\\July 1995}
\pagebreak

\section*{Introduction}
Magnetic fields permeate most astrophysical systems, but their origin is
unknown. The presence of such large-scale magnetic fields can affect processes
like the formation of galaxies\cite{galform} and
nucleosynthesis\cite{nucsynth}.
The existence of a galactic magnetic field, of order $10^{-6}$G on a 100kpc
scale, has been known for some time, but its origin is disputed. One theory
generates it by means of amplifying a `primordial' seed field as small as
$10^{-20}$G\footnote{Recently Plaga\cite{nature}
has devised a method to observe and
measure magnetic fields as weak as $10^{-24}$G in the inter-galactic voids,
and so it may soon be possible to measure the primordial field directly.}
via a galactic
dynamo mechanism\cite{dynamo}, but this leaves the question of where the
seed field came
from? There are a number of scenarios:
by quantum fluctuations in an inflationary phase\cite{infl},
by fluctuations inherent in the plasma\cite{timray} or
as a result
of a cosmological phase transition\cite{otherpt}, such as that thought to
decouple the
electromagnetic and weak forces\cite{vach}\cite{enqol}, or occur during
inflation\cite{kostas}. Here we consider the primordial
magnetic field formed during the electro-weak phase transition.

To facilitate the analysis it is useful to employ the definition for the
electromagnetic field strength introduced by 't Hooft\cite{tHooft}
when studying magnetic monopoles.\footnote{The original motivation for this
was as a way of giving monopoles a magnetic field.}
This collapses to the standard definition if the Higgs field is taken to
condense with the same phase throughout space. The important idea is that if
the Higgs field possesses different phases in neighbouring regions of space
then the
resulting gradients in phase between these regions may produce a magnetic
field. One needs to be careful, however, in case these fields are just a
gauge artefact and can be removed via a suitable transformation.

Following the symmetry breaking produced by a phase transition
the phase of the Higgs field is expected to vary randomly from horizon volume
to horizon volume, through arguments of causality. In practice, space may be
uncorrelated on a shorter scale, corresponding to the
thermal correlation length, given by the inverse of the vector boson
mass, $m_{V}$.
Vachaspati\cite{vach} has pointed out that this will give rise to
phase gradients between these correlation volumes,
which, using 't Hooft's definition of the field strength, may in turn
give rise to a magnetic field. What is more, for causally disconnected
regions it is impossible to gauge these phase differences away. Thus, it is
possible that a cosmological phase transition may well produce a physical
magnetic field. What is an open question, however, is how persistent this
field will be? Thermal fluctuations immediately following the phase
transition and, later on, magnetohydrodynamical effects may both adversely
affect the strength of the field. Here we shall consider only the effect
of the fluctuations on the field strength since if this is too detrimental
then there will be nothing left for the magnetohydrodynamics to work on.
Such thermal fluctuations have not been taken into account in previous
analyses.

In the first section we review 't Hooft's field strength definition
in the context of the Weinberg-Salam model, and discuss
previous attempts to estimate the root-mean-square field by means of
statistical analyses. Previous work\cite{vach}\cite{enqol}
only considered a second order
phase transition. Here we also discuss the relative strength of the field
produced by an equivalent first order transition.
In the second part we analyse how stable such a field is
to thermal fluctuations in the period immediately following a phase transition,
the final section containing our conclusions and suggestions for future work.

\section*{Field Strength}

In the Weinberg-Salam model, in which
SU(2)$\times$U(1)$_{Y}$ gets broken down to U(1)$_{em}$ by the usual Higgs
scalar, $\phi$, the electromagnetic field strength can be written in the
form\cite{tHooft}
\begin{eqnarray}
  F^{em}_{\mu\nu} & \equiv &
  \sin{(\theta_{W})}n^{a}F^{a}_{\mu\nu}+\cos{(\theta_{W})}F^{Y}_{\mu\nu}
  \nonumber\\
   & &
   -i4g^{-1}\eta^{-2}\sin{(\theta_{W})}
   [(D_{\mu}\phi)^{\dagger}D_{\nu}\phi-(D_{\nu}\phi)^{\dagger}D_{\mu}\phi]
   \nonumber
\end{eqnarray}
where
$D_{\mu}=\partial_{\mu}-ieA^{em}_{\mu}$,
$A^{em}_{\mu}=\sin{(\theta_{W})}n^{a}W^{a}_{\mu}+\cos{(\theta_{W})}B_{\mu}$
and
$n^{a}=-2\phi^{\dagger}\sigma^{a}\phi/\eta^{2}$.
A straightforward simplification of the right-hand-side yields
\begin{eqnarray}
  F^{em}_{\mu\nu} & = &
  \partial_{\mu}A^{em}_{\nu}-
  \partial_{\nu}A^{em}_{\mu}-
  i4g^{-1}\eta^{-2}\sin{(\theta_{W})}
  [(\partial_{\mu}\phi)^{\dagger}\partial_{\nu}\phi-
  (\partial_{\nu}\phi)^{\dagger}\partial_{\mu}\phi]
\label{boing}
\end{eqnarray}
which, it is seen, collapses to the usual form of the field strength
if $n^{a}=(0,0,1)$ (corresponding to $\phi=(0,\eta)$).

As mentioned earlier, Vachaspati\cite{vach} suggested that gradients in the
phase between causally
disconnected regions can produce an electromagnetic field. Note that we can
perform a gauge transformation on (\ref{boing}) to set the gauge
fields to zero leaving all the information contained in the Higgs terms.
Thus it is seen that it is possible to have a non-zero magnetic field even if
the gauge fields are zero.

Setting $\phi=\exp{(iT)}\rho$ and $\rho=|\rho|\underline{l}$, where
$T$ is an element of the Lie algebra of SU(2)$\times$U(1) and
$\underline{l}$ is a constant, unit, complex two-vector,
\begin{eqnarray}
  F^{em}_{\mu\nu} & = &
  -i4g^{-1}\eta^{-2}\sin{\theta_{W}}\left[\rho^{\dagger}
  \left((\partial_{\mu}e^{-iT})(\partial_{\nu}e^{iT})-
  (\partial_{\nu}e^{-iT})(\partial_{\mu}e^{iT})\right)\rho\right.\nonumber\\
   & &
   \left.+2\rho^{\dagger}
  (\partial_{\mu}e^{-iT})e^{iT}\partial_{\nu}\rho
  -2\rho^{\dagger}
  (\partial_{\nu}e^{-iT})e^{iT}\partial_{\mu}\rho\right].
\end{eqnarray}
Since we expect the modulus of the scalar field to be uniform across space, to
leading order, during a second order phase transition,
the last two terms in the bracket will give
non-zero contributions to $F_{0i}$ only and hence not contribute to
the magnetic field. They will produce an electric
field however. This
field is shortlived as it is driven by the rising $|\rho|$ and will
drop to zero as $|\rho|\rightarrow\eta$, the global minimum of the
effective potential, as the transition ends. A question
worth considering is whether the magnetic field produced by the associated
current (flowing across the boundary between uncorrelated domains) has
sufficient time to freeze in on relevant scales. We mention several points
concerning this electric field
during our analysis, but concern ourselves, for the most part,
with a study of the magnetic
component of the field tensor, and leave a detailed study of the effect of the
transitive electric field aside for future work. In any case, the magnetic
field induced by the transitive electric field is unlikely to be larger
than the one considered here.

By taking the phase to perform a random walk of fixed step length along a line
passing through $N$ correlation volumes, Vachaspati estimated the gradient
along the line to be $\partial_{\mu}\phi\sim \eta/(\sqrt{N}\xi)$ where $\eta$
is the scale of breaking and $\xi\simeq m_{V}^{-1}$
is the correlation length at formation.
On large scales the intergalactic plasma has a very large conductivity and
the flux is frozen in to the co-moving vacuum. Using this
we can estimate the size
of the field at
time $t$ as $B_{N}\sim gT^{2}/4N$ where $g$ is the SU(2) gauge coupling.
On a 100kpc scale this gives a field today of magnitude $\sim 10^{-30}$G,
much too small to give rise to the observed galactic field.
Other statistical analyses are possible however. By assuming that
the magnetic fields in neighbouring domains are uncorrelated Enqvist
and Olesen\cite{enqol}
have managed to obtain a root-mean-square field strength of
$B_{rms}\propto N^{-\frac{1}{2}}$ giving a primordial field today of
$\sim10^{-18}$G which is much more promising. Since this is a line average,
and the galactic magnetic field is observed by Faraday rotation, this may
well be a more realistic estimate. However, the question of
whether it is legitimate to regard the flux in neighbouring cells as
uncorrelated is still unanswered. For example, if we consider estimates
calculated in
a $N\times N\times N$ volume by taking the na\"{\i}ve model of the phase
gradients performing a random walk through the $N^{3}$ cells,
then we obtain an estimate proportional to $N^{-\frac{3}{2}}$
which is far too small. Clearly considerable work is still required
before we can be sure that the estimates we are making, based on this
mechanism for producing the field, are accurate.

Another feature of these estimates is that they assume a spinodal
decomposition, and so, for the period we shall consider (and
afterwards), there are no large regions of false vacuum. The
magnetic field lies entirely within the true vacuum, along the
boundaries between uncorrelated domains. In the case of a
first order transition things are not so clear cut. Here the
transition proceeds via tunnelling, bubbles of true vacuum nucleating
in the false one, then expanding rapidly, colliding and coalescing
with other bubbles of true vacuum until they fill the Universe
and the transition is complete.

Each time two bubbles collide, a ring of magnetic field is generated
around the region of intersection. If the bubble radius on collision
is of order $\xi$, then $\xi$ is usually\cite{kibbvil} greater
than the thermal correlation length\footnote{There is a need for some
  caution here as for a slow transition some bubbles may nucleate
  at a lower temperature than others but for most cases this will be
  negligible.}, and thus the phase gradient within
two collided bubbles will in general be less than that within two
neighbouring, uncorrelated domains in the equivalent second order phase
transition. Because of this we would expect the magnetic field frozen
in on
large scales to be weaker than that produced by a second order
transition.
It is possible to  quantify this slightly better.

As a crude estimate $\xi\simeq(v/\gamma)^\frac{1}{4}$ where $v$ is the
speed of the bubble wall and $\gamma$ is the bubble nucleation rate
per unit time per unit volume. For the electro-weak transition $v\simeq
0.1-1$ \cite{kibbvil} whilst $\gamma$ is related to the height of the
barrier between the true and false vacua and hence the strength of the
transition. Roughly speaking, the ``stronger'' the transition,
the higher the barrier and, consequently, the lower the rate.

\section*{Stability of the Field}

Having estimated the strength of the field that
may be produced by a cosmological
phase transition, we now turn to the question of how stable this field is to
thermal fluctuations in the period
immediately following its creation. To do this we adapt a
method developed\cite{davrob}
to study the formation of topological defects. We restrict ourselves to the
case of a second order transition. The field produced by a first order phase
transition is likely to be more stable to thermal fluctuations.

 From the electroweak Langrangian we can derive the equation of motion
for the Higgs field,
\begin{eqnarray}
  (\partial_{\mu}+ieA^{em}_{\mu})
  (\partial^{\mu}-ieA^{em\mu})\phi
  +\frac{\partial V}{\partial|\phi|^{2}}\phi & = & 0.
\end{eqnarray}
In \cite{davrob} it was shown that the effect of gauge fields was small, and
in fact only helped the stability of any fields present. Also, as
mentioned above, we can make a transformation to move all the magnetic field
information into the Higgs gradients. For these reasons it is
justifiable to ignore the gauge fields.

As in the previous section,
we make the substitution $\phi=\exp{(iT)}\rho$ where
here, without loss of generality, we take $\rho=|\rho|(0,1)$.
This yields the following equation for $\rho$ and $T$:
\begin{eqnarray}
  \partial_{\mu}\partial^{\mu}\rho
  +2e^{-iT}(\partial_{\mu}e^{iT})\partial^{\mu}\rho
  +e^{-iT}(\partial_{\mu}\partial^{\mu}e^{iT})\rho
  +\frac{\partial V}{\partial \rho^{2}}\rho
  & = & 0.
\end{eqnarray}
To make further progress it is useful to make a couple of
assumptions. Taking $T=\alpha\underline{m}.\underline{\sigma}+\beta$
for SU(2)$\times$U(1) the first assumption is that
$\alpha$ is constant and small,
whilst the second is that we can ignore the
U(1) symmetry (identical to assuming constant $\beta$). Thus we are
constraining the degree to which the phase of the Higgs can vary. Our
justification for doing this is partly practical and partly motivated
by the fact that if we are to let any one part of the phase vary it has to
be $\underline{m}$. If this is constant, then the resulting
magnetic field is always zero. This is not true in the case of
constant $\alpha$ and $\beta$ - $\beta$ in fact only affects the
electric field. We have simply taken the minimal
requirements to obtain a non-zero magnetic field.

Writing $\underline{\alpha}=\alpha\underline{m}$ we
obtain
\begin{eqnarray}
  \partial_{\mu}\partial^{\mu}\rho
  -(\partial_{\mu}\underline{\alpha}.\partial^{\mu}\underline{\alpha})\rho
  +\frac{\partial V}{\partial \rho^{2}}\rho
  & = & 0\nonumber\\
  \partial_{\mu}\partial^{\mu}\underline{\alpha}
  +2\partial_{\mu}\underline{\alpha}\partial^{\mu}\rho/\rho
  & = & 0
\end{eqnarray}
where $\rho$ now denotes $|\rho|$.
As expected, the equations for $\alpha_{1}$, $\alpha_{2}$ and $\alpha_{3}$ are
identical - there is no preferred breaking direction. These equations
are very similar to those derived in\cite{davrob} and so we make heavy use
of their perturbation analysis. To avoid repitition, full details can be
obtained there, and here we sketch only the necessary detail for our analysis.

Consider the perturbed system where
\begin{eqnarray}
        \rho=\rho_{B}+\rho_{F} & , &
        \underline{\alpha}=
        \underline{\alpha}_{B}+\underline{\alpha}_{F},
\end{eqnarray}
and anything with a subscript $F$ is considered small.
We have the following equations of motion:
\begin{eqnarray}
        \partial^{2}(\rho_{B}+\rho_{F})-
        \big[
        \partial_{\mu}
        (\underline{\alpha}_{B}+\underline{\alpha}_{F}).
        \partial^{\mu}
        (\underline{\alpha}_{B}+\underline{\alpha}_{F})
        \big]
        (\rho_{B}+\rho_{F}) & & \nonumber\\
        +(\rho_{B}+\rho_{F})\left[
        \frac{\partial V}{\partial\rho^{2}}+
        \rho_{F}\frac{\partial}{\partial\rho}
        \left(\frac{\partial V}{\partial\rho^{2}}\right)
        \right] & = & 0,\nonumber\\
        \partial^{2}(\underline{\alpha}_{B}+\underline{\alpha}_{F})
        +2\partial_{\mu}(\underline{\alpha}_{B}+\underline{\alpha}_{F})
        \partial^{\mu}(\rho_{B}+\rho_{F})
        /(\rho_{B}+\rho_{F})
         & = & 0.\nonumber
\end{eqnarray}
Taking the zeroth-order in the perturbation expansion we obtain the
background equation
\begin{eqnarray}
        \partial^{2}\rho_{B}-
        (\partial_{\mu}\underline{\alpha}_{B}.
        \partial^{\mu}\underline{\alpha}_{B})\rho_{B}+
        \frac{\partial V(\rho)}{\partial\rho^{2}}\rho_{B}
         & = & 0.
\end{eqnarray}
Making use of the finite temperature form of the potential for a
second order phase transition,
\begin{eqnarray}
        V(\phi,T) & = &
        \frac{1}{4}\lambda|\phi|^{4}-
        \ha(\lambda\eta^{2}-\tilde{\lambda}T^{2})|\phi |^{2}+
        \frac{1}{4}\lambda\eta^{4},
\end{eqnarray}
but ignoring the $T^{2}$ term, since initially
$T=T_{G}<T_{C}=\sqrt{\lambda/\tilde{\lambda}}\eta$ and $T$ decreases with time,

\begin{eqnarray}
        \partial^{2}\rho_{B}-
        (\partial_{\mu}\underline{\alpha}_{B}.
        \partial^{\mu}\underline{\alpha}_{B})\rho_{B}-
        \ha\lambda\eta^{2}\rho_{B}
         & = & 0.
\end{eqnarray}
If we assume that in the initial configuration
$\underline{\alpha}_{B}$ varies on some spatial scale $1/k$, this
implies that
\begin{eqnarray}
        \partial_{t}^{2}\rho_{B}= m_{I}^{2}\rho_{B}
        \equiv (\ha\lambda\eta^{2}-k^{2}\underline{\alpha}^{2}_{B})\rho_{B} & &
\end{eqnarray}
since we can take $\underline{\alpha}_{B}$
to be independent of time. We see that
there is an exponential instability if
$k^{2}\underline{\alpha}_{B}^{2}\leq\ha\lambda\eta^{2}\equiv k_{C}^{2}$,
corresponding to the modes
that drive the transition at early times. Hence, for $k\alpha_{B}<k_{C}$,
\begin{eqnarray}
        \rho_{B} & \simeq &
        {\cal{A}}_{B}\exp{(m_{I}(t-t_{G}))}
\end{eqnarray}
where $m_{I}^{2}\simeq k_{C}^{2}$.

Note that taking the
initial configuration to fluctuate on a scale $1/k_{C}$, gives
$\partial_{i}g\simeq k_{C}$, whilst, from above,
$\dot{\rho_{B}}/\rho_{B}\simeq k_{C}$. This means that, for this period
at least, the strength of the electric field will be of the same order of
magnitude as the magnetic one.

Taking the assumption that the length-scale of fluctuations in
$\underline{\alpha}_{F}$
is the same as that in
$\rho_{F}$
(we later demonstrate that this is self-consistent) we now show that
$\rho_{F}/\rho_{B}<1$.

Consider first-order perturbations to the equation of motion:
\begin{eqnarray}
        \partial^{2}\rho_{F}-
        2(\partial_{\mu}\underline{\alpha}_{B}).
        (\partial^{\mu}\underline{\alpha}_{F})\rho_{B}-
        (\partial_{\mu}\underline{\alpha}_{B}).
        (\partial^{\mu}\underline{\alpha}_{B})\rho_{F}-
        \ha\lambda\eta^{2}\rho_{F}+
        \lambda\rho_{B}^{2}\rho_{F}
         & = &
        -g\psi^{2}\rho_{B}
        \label{pig}
\end{eqnarray}
where we have introduced a thermal
noise term by coupling the scalar, $\phi$, to
another, $\psi$, in thermal equilibrium.

Assuming that the wave number of fluctuations in $\alpha_{F}$ is less
than $k_{C}$ so that the second term in (\ref{pig}) does not dominate,
\begin{eqnarray}
  (\partial _{t}^{2}-\nabla^{2})\rho_{F}
  -m_{I}^{2}\rho_{F}
  \simeq -g\psi^{2}\rho_{B}
\end{eqnarray}
which can be solved by Green's function methods:
\begin{eqnarray}
  \rho_{F}(\underline{x},t)=
  -g\int^{t}_{0}d\tau d^{3}y
  G_{ret}(t-\tau,\underline{x}-\underline{y})\psi^{2}(\tau,\underline{y})
  \rho_{B}(\tau,\underline{y})
\end{eqnarray}
where
\begin{eqnarray}
  G_{ret}(x)=
  -\frac{1}{(2\pi)^4}\int d^{4}p
  \frac{\exp{(-ipx)}}{(p_{0}+i\epsilon)^{2}-\underline{p}^{2}+m_{I}^{2}}.
\end{eqnarray}
Now, for a rapid phase transition we can ignore the time-dependence of
$\psi^{2}$. Substituting in the lower bound for $\rho_{B}$, we obtain
\begin{eqnarray}
  \rho_{F}(t,\underline{x})\leq
  \frac{g{\cal{A}}_{0}}{2m_{I}^{2}}\psi^{2}\exp{(m_{I}t)}.
\end{eqnarray}
If we take $\psi$ to be a self-interacting scalar field with self-coupling of
order one, then in thermal equilibrium $\psi^{2}\sim T^{2}$ and so
\begin{eqnarray}
  \rho_{F}/\rho_{B}\leq gT^{2}/m_{I}^{2}.
\end{eqnarray}
Provided that $\tilde{\lambda}>g$, initially
$T=T_{G}<T_{C}=\sqrt{\lambda/\tilde{\lambda}}\eta\leq\sqrt{\lambda/g}\eta$
 and so
\begin{eqnarray}
  \rho_{F}/\rho_{B}<1.
\end{eqnarray}
As an aside,
it is straightforward to show that
\begin{eqnarray}
  \partial_{0}\rho_{F}/\partial_{0}\rho_{B}\simeq
  \rho_{F}/\rho_{B}-
  g\left(\frac{T}{m_{I}}\right)^{3}\left(\frac{T}{m_{Pl}}\right)
\end{eqnarray}
and so $\partial_{0}\rho_{F}/\partial_{0}\rho_{B}<1$.
Since $\partial_{0}\rho/\rho\simeq
\partial_{0}\rho_{B}/\rho_{B}+
\partial_{0}\rho_{F}/\rho_{B}$
this suggests that, for its short lifetime, fluctuations have no major impact
on the
magnitude of the electric field.

For self-consistency, however, we still have to show that
$\rho_{F}/\rho_{B}<1$ implies $\partial_{i}\alpha_{F}<1/k_{C}$.
Our equation for $\underline{\alpha}_{F}$ is
\begin{eqnarray}
  \partial^{2}\underline{\alpha}_{F}
  -2(\underline{k}_{F}.\underline{k}_{C}) \underline{\alpha}_{F}
   & = &
   2(\underline{k}_{C}.\underline{k}_{F})(\rho_{F}/\rho_{B})
   \underline{\alpha}_{B}
\end{eqnarray}
Note that only $\underline{\alpha}_{F}$
and $\rho_{F}$ are functions of $\underline{k}_{F}$. It is
easy to Fourier transform and solve this using the Green's function
method once more. Provided that $k_{F}\gg k_{C}$, we find that
\begin{eqnarray}
  \tilde{\underline{\alpha}}_{F}(\underline{k}_{F},\tau)\sim
  \frac{\underline{k}_{C}.\underline{k_{F}}}{k_{F}^2}
  \frac{\tilde{\rho_{F}}}{\rho_{B}}
  \underline{\alpha}_{B}
\end{eqnarray}
where a tilde denotes a Fourier transform, and so short wavelengths
are suppressed compared to long wavelength
inhomogeneities. Thus, thermal fluctuations do not destroy the generated
magnetic field.

\section*{Conclusions}

We have seen then that magnetic fields produced by gradients in the Higgs
field between uncorrelated domains following a second order phase transition
are stable
to thermal fluctuations, despite not being of a topological
nature. Whether such fields can then survive the ensuing
microhydrodynamical processes to achieve the required magnitude
today is a question that still needs to be answered, and one which
depends heavily upon correct estimates of the initial strength of the
field. As shown, there is still some doubt about which of the current
estimates, if any, is the correct one.

The case of a first order phase transition is even worse
understood. We have seen how we might expect the field produced to be
generally smaller in magnitude than one produced in an equivalent
second order transition. For a {\em weakly} first order transition we
might expect the formation and stability to differ little from the
second order case. A detailed analysis is yet to be done however.

Another feature that we have observed is the transitive electric field
produced as the scalar field drops into the minimum of the
potential. Although, at least for part of its brief life, it is of the same
strength as the magnetic field, whether
it survives for long enough to imprint an induced magnetic field on the
background is still an open question. However, the magnitude
of the induced field is unlikely to be larger than the magnetic field
considered here, and the same stability analysis applies.

More generally, although we take here the specific
case of the electro-magnetic field, produced by
the electro-weak phase transition, it should be noted that using the 't Hooft
definition, magnetic fields are possible remnants of any phase transition in
which a non-Abelian symmetry is broken, the field being associated with the
unbroken Abelian\footnote{Their non-Abelian counterparts acquire a
magnetic mass and are consequently screened by the plasma.}
generators. Hence magnetic fields may exist associated with
hypercharge, for example.

\section*{Acknowledgements}

This work is supported in part by PPARC and EPSRC. We would like to
thank Tim Evans for discussions.

\end{document}